\documentclass[useAMS, usenatbib]{mn2e}
\usepackage{graphicx, rotate}
\usepackage{natbib, lscape}
\usepackage{latexsym, amssymb, verbatim}

\title[Molecules as heating probe in Perseus]{First observations
of CN(2-1), HCO$^{+}$(3-2) and C$_{2}$H(3-2) emission lines in the
Perseus cluster: constraints on heating mechanisms in the cluster
gas.} \author[Bayet et al.]{E. Bayet$^{1}$\thanks{E-mail:
eb@star.ucl.ac.uk; daw@star.ucl.ac.uk; twh@ast.leeds.ac.uk;
sv@star.ucl.ac.uk}; S. Viti$^{2}$, T. W.
Hartquist$^{3}$ and D.A. Williams$^{2}$\\
$^{1}$Sub-Department of Astrophysics, University of Oxford, Denys
Wilkinson Building, Keble Road, Oxford OX1 3RH\\
$^{2}$Department of Physics and Astronomy, University
College London, Gower Street, London WC1E 6BT, UK\\
$^{3}$School of Physics and Astronomy, University of Leeds, Leeds
LS2 9JT, UK\\}

\begin{document}

\date{Accepted ; Received ; in original form }

\pagerange{\pageref{firstpage}--\pageref{lastpage}} \pubyear{2010}

\maketitle

\label{firstpage}

\begin{abstract}
We present the first observations of emission lines of CN(2-1),
HCO$^{+}$(3-2) and C$_{2}$H(3-2) in the Perseus cluster. We
observed at two positions: directly at the central galaxy, NGC
1275 and also at a position about 20$''$ to the east where
associated filamentary structure has been shown to have strong CO
emission. Clear detections in CN and HCO$^{+}$ transitions and a
weak detection of the C$_{2}$H transition were made towards NGC
1275, while weak detections of CN and HCO$^{+}$ were made towards
the eastern filamentary structure. Crude estimates of the column
densities and fractional abundances (mostly upper limits) as
functions of an unknown rotational temperature were made to both
sources. These observational data were compared with the outputs
of thermal/chemical models previously published by \citet{Baye10c}
in an attempt to constrain the heating mechanisms in cluster gas.
We find that models in which heating is dominated by cosmic rays
can account for the molecular observations. This conclusion is
consistent with that of \citet{Ferl09} in their study of gas
traced by optical and infrared radiation. The cosmic ray heating
rate in the regions probed by molecular emissions is required to
be at least two orders of magnitude larger than that in the Milky
Way.
\end{abstract}


\section{Introduction}\label{sec:intro}

The Perseus Cluster is one of the nearest galaxy clusters and is
the brightest X-ray cluster in the sky. The cluster and its
central galaxy NGC 1275 have been the focus of intense study for
many years, at X-ray, optical, IR and millimetre wavelengths. The
first molecular detections were of CO rotational emission towards
the centre of NGC 1275 \citep{Laza89, Mira89, Reut93, Brai95,
Inou96, Brid98, Lim08}. Observations have now demonstrated that CO
emission also extends to some tens of kpc from the central galaxy
\citep{Salo06,Salo08a,Salo08b,Salo11} and is strongly correlated
with the filamentary structure observed in H$\alpha$ \citep{Hu83,
Cowi83,Cons01} within the hot gas detected in X-rays at 0.5 keV
\citep{Fabi08}. The H$\alpha$ structure is also correlated with
warm H$_{2}$ emission \citep{Edge02,Wilm02}. Some of the molecular
gas towards the cluster must be at high density ($\geqslant
10^{4}$ cm$^{-3}$). \citet{Salo08a} have made the first detection
of the high density tracer HCN(3-2) emission towards the centre of
NGC 1275.

The nature of the filamentary structure is the subject of current
discussion. \citet{Salo11} note two possibilities: either that the
CO filaments form far from the galaxy's centre from uplifted warm
gas, eventually falling back \citep{Reva08}, or that the molecular
gas is entrained and dragged out of the galaxy by rising hot gas.
Evidently, further observations of molecular gas may help to
determine its origin. In this paper, we present observations of
emission lines of CN(2-1), HCO$^{+}$(3-2) and C$_{2}$H(3-2)
towards the central galaxy and also towards a position 20$''$ to
the east where there is strong CO emission in the associated
filamentary structure.

These three species were identified in a theoretical study
\citep{Baye10c} as tracers of regions influenced by the
dissipation of turbulence and waves, heating the gas and
accelerating cosmic rays (see also \citealt{Craw92, Pope08a}).
\citet{Ferl09} developed models of the optical and infrared
emission filamentary regions in order to infer the rate at which
mechanical energy is dissipated, and to determine the cosmic ray
background produced by the interaction. \citet{Ferl09} concluded
that the heating rate per hydrogen nucleus due to dissipation or
cosmic ray induced ionisation must be 300 times higher in the
optical emission filaments than in the local interstellar medium.

In their work, \citep{Baye10c} identified the species CN, C$_{2}$H
and HCO$^{+}$ as sensitive to changes in cosmic ray ionisation
rate or to an additional source of heating such as dissipation.
\citet{Meij11} came to a similar conclusion. Here, we present the
observational follow-up to that theoretical study; we aim to
determine whether dissipation or cosmic ray induced ionisation
dominates the heating.

The paper is organised as follows: in Section \ref{sec:obs} we
present the first observations of CN(2-1), C$_{2}$H(3-2) and
HCO$^{+}$(3-2) emissions in both NGC 1275 and a position 20$''$ to
the east where emission in the CO filamentary structure is known
to be strong (see Figure \ref{fig:0}). In Section \ref{sec:resu}
we analyse the data and provide rough estimates of column
densities and fractional abundances of the three molecules in NGC
1275 and in the filamentary region. Section \ref{sec:comp}
compares the results of the observations with the models of
\citet{Baye10c} to try to identify whether cosmic ray heating or
dissipation is the dominant heating mechanism, or whether a
combination of both is required. Section \ref{sec:con} discusses
the results and gives our conclusions.

\section{Observations}\label{sec:obs}

\begin{figure}
    \centering\includegraphics[trim=1mm 0mm 0mm 0mm,clip, width=8cm]{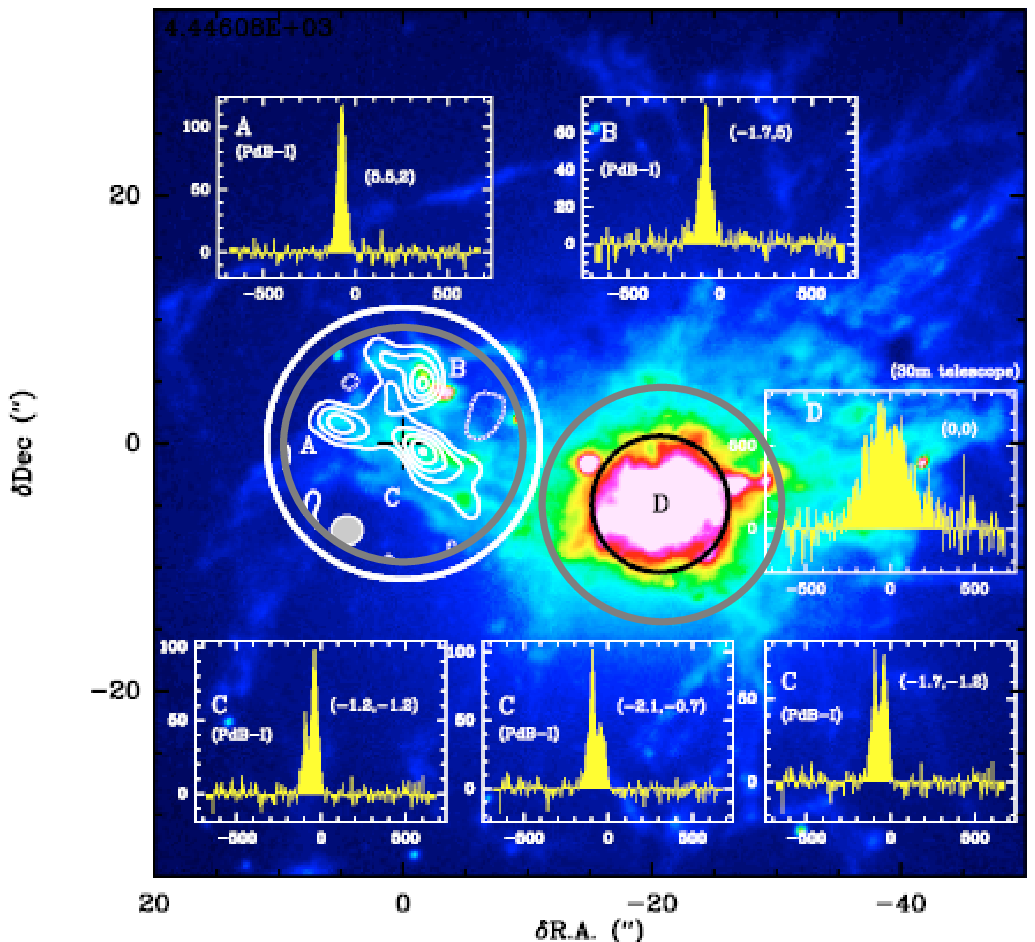}
    \caption{Extracted from \citet{Salo08b}. This figure presents the positions and size of our
    JCMT beams (20$''$ grey circles) used for our observations relative to the integrated
    CO(2-1) intensity (white contours) seen on an H$\alpha$ image from \citet{Cons01} (colour
    scale) of NGC 1275. The white circle shows the Plateau de Bure
    Interferometer primary beam of the \citet{Salo08b} CO data. The black
    circle located on NGC 1275 (at (-20$''$,-5$''$)
    offsets from the centre on the map) corresponds to the 11.7$''$ beam of the IRAM-30m
    single-dish telescope CO data from \citet{Salo06}.}\label{fig:0}
\end{figure}

The observations were performed using the James Clerk Maxwell
Telescope (JCMT) for the detection of the CN(2-1) line ($\nu=$
226.874 GHz), C$_{2}$H(3-2) transition ($\nu=$ 262.004 GHz) and
HCO$^{+}$(3-2) line ($\nu=$ 267.557 GHz). In this first attempt of
detection, we did not seek to resolve the fine structure line
emission of the CN nor of the C$_{2}$H but rather observe the
emission coming from the whole (2-1) and (3-2) group,
respectively. Every two or three hours the pointing, focus and
calibration were performed carefully on planets (Mars and Jupiter)
and on evolved stars. The pointing error was estimated to be
$\leq$ 3$''$.

\begin{figure}
    \includegraphics[width=8cm]{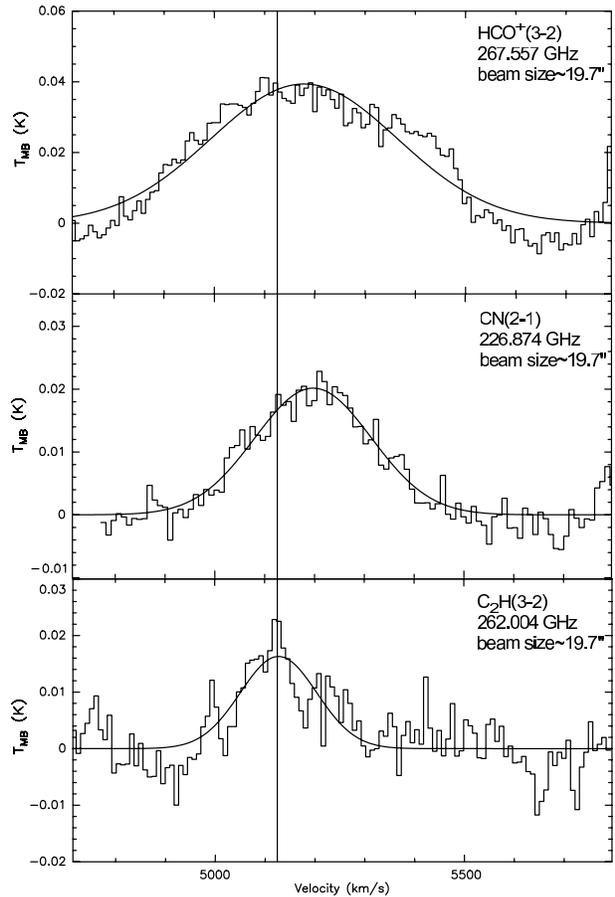}
    \caption{Molecular spectra obtained in NGC 1275, galaxy located at the
    centre of the Perseus cluster. \textit{Top panel: }HCO$^{+}$(3-2) line,
    \textit{Middle panel: }CN(2-1) transition and \textit{Bottom panel: }C$_{2}$H(3-2) line.
    The vertical black line represents the position of the $V_{lsr}$=5125 km s$^{-1}$
    for the CO gas in NGC 1275 (see \citealt{Brid98}). In each panel, the back curve shows the result of the Gaussian fit profile
    applied whose output parameters are listed in Table \ref{tab:obs} (see Section
    \ref{sec:obs}). Despite the fact that a Gaussian profile can be fitted to the
    C$_{2}$H(3-2) spectrum, this observation can not be considered as a detection since
    its corresponding signal-to-noise ratio is below the usual extragalactic cutoff of 3.}\label{fig:1}
\end{figure}

\begin{figure}
    \includegraphics[width=8cm]{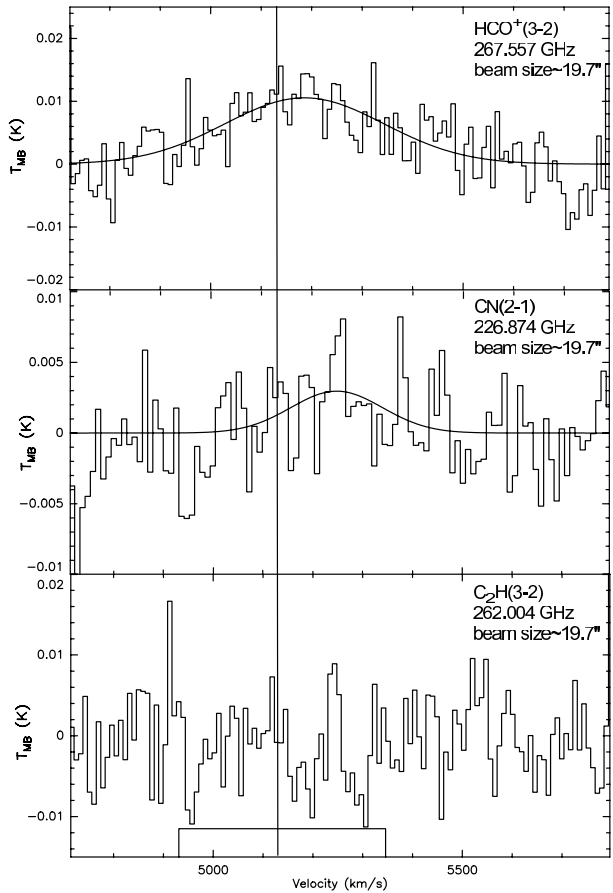}
    \caption{Molecular spectra obtained in a shifted position to the
    centre of the Perseus cluster corresponding to a filamentary structure.
    \textit{Top panel: }HCO$^{+}$(3-2) line, \textit{Middle panel: }CN(2-1)
    transition and \textit{Bottom panel: }C$_{2}$H(3-2) line. See caption of
    Figure \ref{fig:1}. All the observations presented in this figures have to
    be considered as upper limits only, even if a Gaussian line profile can been seen for
    the HCO$^{+}$(3-2) and the CN(2-1) lines. Indeed, for these three observations,
    the signal-to-noise ratio obtained is actually below the usual extragalactic cutoff of 3.
    Any Gaussian fit seen has thus to be considered as only indicative of a
    \emph{potential} detection. The velocity window seen on the spectrum of the
    C$_{2}$H(3-2) line is indicative too, for marking the expected position of the
    line, if it were detected.}\label{fig:2}
\end{figure}

The JCMT observations were made between August and December 2010
under medium weather conditions ($\tau_{225} \approx$ 0.08-0.15).
We employed a beam switch mode with a throw of 95$''$ and used the
RxA3 receiver on JCMT for detecting these three molecular
emissions, coupled with the ACSIS digital autocorrelation
spectrometer with a bandwidth of 1000MHz because the lines were
expected to be broad (see Figures \ref{fig:0} and \ref{fig:2}).
The HPBW and the main beam efficiencies of the JCMT at $\nu \sim$
250 GHz are 20$''$ and 0.69\footnote{See JCMT website:
http://www.jach.hawaii.edu/JCMT/ instruments/}, respectively (see
the projected JCMT beam on Figure \ref{fig:0} as compared to the
one corresponding to the published data from IRAM-30m and
IRAM-Plateau de Bure Interferometer). The system temperatures
ranged between 200 K and 450 K, depending on the source (either
NGC 1275 or the filament) and on the wavelength. The data
pre-reduction was done using Starlink software (KAPPA, SMURF and
STLCONVERT packages) and subsequently translated to CLASS format
for final reduction. The reduced spectra for NGC 1275 and the
filamentary region are seen respectively in Figures \ref{fig:1}
and \ref{fig:2} and their resulting Gaussian fitting parameters
are displayed in Table \ref{tab:obs}.

The C$_{2}$H(3-2) line in NGC 1275 and the three lines observed in
the filamentary region show a signal-to-noise ratio below the
usual extragalactic cutoff of 3$\sigma$. For clear detections in
the filamentary region, one would require more on-source
integration time. These observations are thus to be considered all
as upper limits, even though in the case of the C$_{2}$H(3-2) line
in NGC 1275 and of the HCO$^{+}$(3-2) and CN(2-1) transitions in
the filament, Gaussian line profiles could be fitted to the
observations (see Figures \ref{fig:1} and \ref{fig:2}). Where fits
could be obtained, the fitting parameters are shown in Table
\ref{tab:obs}. However, note that for the lines of low
signal-to-noise the values should be regarded, with caution, as
indicative; these values are shown in italic in Table
\ref{tab:obs}.

\begin{table*}
  \caption{Observational parameters and Gaussian fits parameters obtained for
  the data set. When the data are upper limits and we have not succeeded in
  fitting any Gaussian line profiles, we have used a symbol ``$-$'' in the table
  and given for the T$_{peak}$ value a 3 $\sigma$ estimate based on the value
  of the observed rms. When fitting values are seen in a italic text
  font, it means that the observations are also to be considered as upper
  limits but it has been possible to fit a Gaussian line profile to the data
  (see text in Section \ref{sec:obs}).}\label{tab:obs}
  \resizebox{11cm}{!}{
  \begin{tabular}{l c c}
  \hline
  & NGC 1275 & Filament\\
  RA(J2000) (h:m:s)& 03:19:48.20 & 03:19:50.00 \\
  DEC(J2000) ($^{\circ}$ : ' : '')& 41:30:42.0 & 41:30:47.0\\
  Position in Figure \ref{fig:0} & D & A, B \& C \\
  \hline
  \textbf{HCO$^{+}$(3-2), $\nu=$ 267.557 GHz} & &\\
  Tsys (K) & 253 & 315\\
  Integration time$^{a}$ (mins) & 20 & 18\\
  $\int$(T$_{mb}$ dv) (K km s$^{-1}$) & 18.3$\pm$0.4 & \emph{4.2$\pm$0.4}\\
  V$_{peak}$ (km s$^{-1}$) & 5178.7$\pm$5.0 & \emph{5185.7$\pm$11.1}\\
  FWHM (km s$^{-1}$) & 437.5$\pm$10.0 & \emph{372.1$\pm$34.2}\\
  T$_{peak}$ (mK) & 39.4 & \emph{10.5}\\
  rms (mk)& 7.9 & 8.6\\
  \hline
  \textbf{CN(2-1), $\nu=$ 226.874 GHz} & &\\
  Tsys (K) & 197 & 252\\
  Integration time$^{a}$ (mins) & 27 & 20\\
  $\int$(T$_{mb}$ dv) (K km s$^{-1}$) & 5.9$\pm$0.2 & \emph{0.7$\pm$0.2}\\
  V$_{peak}$ (km s$^{-1}$) & 5197.3$\pm$4.3 & \emph{5248.9$\pm$33.5}\\
  FWHM (km s$^{-1}$) & 275.9$\pm$9.5 & \emph{214.2$\pm$91.0}\\
  T$_{peak}$ (mK) & 20.2 & \emph{3.0}\\
  rms (mk)& 4.8 & 2.8\\
  \hline
  \textbf{C$_{2}$H(3-2), $\nu=$ 262.004 GHz}& &\\
  Tsys (K) & 306 & 442\\
  Integration time$^{a}$ (mins) & 18 & 18\\
  $\int$(T$_{mb}$ dv) (K km s$^{-1}$) & \emph{3.1$\pm$0.2} & -\\
  V$_{peak}$ (km s$^{-1}$) & \emph{5126.8$\pm$7.4} & -\\
  FWHM (km s$^{-1}$) & \emph{180.8$\pm$17.3} & -\\
  T$_{peak}$ (mK) & \emph{16.3} & $<$23.4\\
  rms (mk)& 8.3 & 7.8\\
  \hline
  \end{tabular}}

  $^{a}$ The integration time listed here is the ON-SOURCE time only.
  \end{table*}

To correct the integrated line intensities for beam dilution
effects, we have assumed optically thin emission (antennae
temperature proportional to the column density in the upper level
of the observed transition) and source sizes of 20$''$ and 11$''$
for NGC 1275 and the filament regions, respectively. These source
sizes have been derived from the CO single-dish and
interferometric maps from \citet{Salo06, Salo08a, Salo08b} (see
also Figure \ref{fig:0}). We have applied this correction before
estimating the column densities and fractional abundances (but
note that the data in Table \ref{tab:obs} do not include any
correction for beam dilution).

We note that the HCO$^{+}$(3-2) line in NGC 1275 is significantly
broader than other lines presented here, and is broader than the
CO(1-0) and CO(2-1) line widths (of about 380 km s$^{-1}$)
reported in \citet{Brid98,Salo06, Salo08a}. The cause of this
broadening is unclear. The range of frequencies covered by the
spectrum seen in Figure \ref{fig:1} (i.e. 267.093-268.036 GHz),
includes only one group of lines potentially strong enough to
blend with the HCO$^{+}$(3-2) emission and cause this broadening:
this is the methanol emission located around 267.5 GHz. However,
since methanol (in any frequency range) has not been detected in
NGC 1275, nor has it been found in any central galaxy of a
cluster, this possible explanation remains hypothetical.

Using previously published Perseus cluster molecular line data, we
have listed in Table \ref{tab:add} some line ratios corrected for
beam dilution effects. We have included additional molecular lines
known to be reliable tracers of relatively high density gas from
\citet{Brid98, Salo08a}. For the filamentary region, we have used
the CO(3-2) integrated line intensity from \citet{Brid98} at the
position closest to ours ((+7$''$;+7$''$) as compared to
(+20$''$;+5$''$)), without finding a better match.

We briefly compare our results for the Perseus Cluster with the
results of other molecular line studies of galaxies.
\citet{Gao04a} give integrated line intensities (K km s$^{-1}$)
for CO in nearby galaxies ranging from a few to a few hundred,
while for HCN the values range from a few tenths to a few tens
(see also \citealt{Aalt02}). Our results (in the same units) for
CN(2-1), C$_{2}$H(3-2) and HCO$^{+}$(3-2) are 5.9, 3.1 and 18.3.
So, in NGC 1275 these three molecular species are typically as
bright as HCN in many other galaxies, though not as bright as CO.

\begin{table}
  \caption{Line ratios using previously published data
  from \citet{Brid98, Salo06, Salo08a} and the
  observations presented here. The line ratios have been calculated only after the
  integrated line intensities of all the observations used here were corrected for beam dilution
  effects using an assumed beam size of 20$''$. When the symbol `-' is used, it means that the data are unknown.
  The CO(3-2) integrated line intensity used in the filamentary calculations from \citet{Brid98}
  does not correspond exactly to the position observed here and therefore
  has to be considered as a lower limit.}\label{tab:add}
  \begin{tabular}{l c c}
  \hline
  & NGC 1275 & Filament\\
  \hline
  HCO$^{+}$(3-2)/CO(3-2) & 2.12 & $<$0.53\\
  HCO$^{+}$(3-2)/HCN(3-2) & 13.46 & - \\
  CN(2-1)/CO(3-2) & 0.68 & $<$0.09\\
  CN(2-1)/HCN(3-2) & 4.31 & - \\
  C$_{2}$H(3-2)/CO(3-2) & 0.36 & -\\
  C$_{2}$H(3-2)/HCN(3-2) & 2.28 & -\\
  \hline
  \end{tabular}
  \end{table}

\section{Column density and fractional abundance estimates}\label{sec:resu}

To convert the integrated line intensities into column densities
and fractional abundances, the hypothesis of a Local
Thermodynamical Equilibrium (LTE) has been assumed such as, for a
species $X$, we have:
\begin{equation}
N_{tot}(X) = \frac{N_{u}}{g_{u}}(X) \times Q(T_{rot}) \times exp
\left( \frac{E_{u}}{k \times T_{rot}} \right)
\end{equation}
where $\frac{N_{u}}{g_{u}}(X)$ is defined by:
\begin{equation}\label{eq:2}
\frac{N_{u}}{g_{u}}(X) = 1.669 \times 10^{17}\times
\frac{\int{T_{mb} dv}_{corr}}{\nu \times (S \mu^{2})}
\end{equation}
where $\int{T_{mb} dv}_{corr}$ corresponds to the integrated line
intensities listed in Table \ref{tab:obs}, corrected from beam
dilution effect, where $Q(T_{rot})$ is the partition function at
the rotational temperature $T_{rot}$, where $E_{u}$ is the energy
of the upper level of the studied transition, where $k$ is the
Planck constant, and where $S \mu^{2}$ is can be found, similarly
to $Q(T_{rot})$ values, in the Cologne Database for Molecular
Spectroscopy (CDMS\footnote{See the website:
http://www.astro.uni-koeln.de/cgi-bin/cdmssearch.}).

Any more sophisticated treatment at this stage of the observations
has been considered as meaningless since only one transition per
molecule has been obtained per source. Since the current dataset
of observations does not allow us to estimate the rotational
temperature $T_{rot}$ i. e. rotational diagrams could not be
constructed, we present in Table \ref{tab:resu} values of the
total column densities for CN, C$_{2}$H and HCO$^{+}$ as functions
of $T_{rot}$. The $T_{rot}$ values listed have been chosen to be
consistent with the range of kinetic temperatures computed
self-consistently in the various models explored by
\citet{Baye10c}.

To estimate the fractional abundance of a molecule from its
estimated total column density, $N_{tot}(X)$ we have assumed the
canonical gas:dust ratio for the Milky Way. This value may not be
appropriate for the Perseus Cluster. We have selected $A_{v}$ to
be equal to either 3 mag or 8 mag. This allows the direct
comparison with results presented in \citet{Baye10c}. Though we
cannot be certain that the material is at either of these visual
extinctions. However, we have selected them as the lower value is
representative of translucent regions and the higher corresponds
to a dark region that does not have such a high column density
that its lifetime would necessarily be significantly limited by
gravitational collapse. We recognize that the whole procedure that
we have used is necessarily very crude, but it does allow rough
estimates of fractional abundances to be made, to compare with
models (see Section \ref{sec:comp}). The derived column densities
and fractional abundances, for both NGC 1275 and the filamentary
region, are presented in Tables \ref{tab:resu} and \ref{tab:abun},
respectively.

\section{Comparison with model predictions: constraints on the origin
of the heating mechanisms in the Perseus cluster}\label{sec:comp}

We have extracted from \citet{Baye10c}, the values of the
fractional abundances of CN, C$_{2}$H and HCO$^{+}$ from five
models that may be appropriate for reproducing the observations
presented here in both NGC 1275 and in the filament region. We
present these values in Table \ref{tab:mod}. They offer five
distinct combinations of cosmic ray ionisation rate (hereafter
$\zeta$ in s$^{-1}$), of additional source of heating, $H$, (in
erg cm$^{-3}$ s$^{-1}$) accounting for the effect of dissipation,
and of FUV radiation field, $I$, (expressed in units of Habing -
see \citealt{Habi68}).

More precisely, Model A has high values of both $H$ and $\zeta$
with a low value of $I$. Model B has low values of $H$, $\zeta$
and $I$. Model C has low values of $H$ and $\zeta$ with a high
value of $I$. Model D has a high value of $H$ and low values of
$\zeta$ and $I$. Model E has a low value of $H$, a high value of
$\zeta$ and a low value of $I$ (See Table \ref{tab:mod}). Each
model explored thermal and chemical properties at two values of
$A_{v}$, i.e. $A_{v}=$3 and 8 mag. These values are intended to
represent conditions near the edge and in the dark interior of a
molecular region.

By comparing directly the observed fractional abundances values
presented in Table \ref{tab:abun} with those modelled listed in
Table \ref{tab:mod}, for the same temperature range (i.e. here, we
compare only observed and modelled fractional abundances at
$T_{rot} \sim T_{K}$), we may be able to identify which heating
process may be present and consistent with the observed emissions
of both NGC 1275 and the filamentary region. A summary of our
conclusions is presented in Table \ref{tab:sum}. As one can see,
there are actually a relatively limited number of models which can
reproduce the observations well. Here we arbitrarily considered
that a model reproduces well an observation when there is a factor
of 5 or less between the modelled and observed fractional
abundances. This factor has been selected such as taking into
account possible uncertainties on the detections (e.g. pointing,
calibration, etc) as well as uncertainties in the gas:dust ratio
that affect the column densities and fractional abundances; also,
the model results for abundances scale with the unknown
metallicity.

Inspection of Table \ref{tab:sum} suggests that the radiation
field has little effect on the resulting chemistry, whatever
values are attained by $H$ and $\zeta$. We shall ignore the FUV
radiation field parameter in the following discussion, and focus
on the effects of varying $H$ and $\zeta$. Note that where the
observations supply only upper limits, we have assumed that the
actual value is close to the upper limit. Further, we note that
the success or failure of the models to match observations appears
to be the same for both the centre, i.e. towards NGC 1275, and the
eastern filamentary position.

Table \ref{tab:sum} shows that individual species may have a match
with observation for several models. For example, CN gives a match
to observations for Models B, E (at $A_{v}=$ 3 mag), Model D and
Model C, while C$_{2}$H could match the observations for Models E
and A. However, if all the molecules are assumed to share the same
space and physical conditions, then we require a single model to
account for all the species considered. Table \ref{tab:sum} shows
that there is such a model, Model E. It can account for CN and
HCO$^{+}$ at $A_{v}=$ 3 mag (though not at 8 mag), and for
C$_{2}$H. The values of $T_{rot}$ associated with all three
molecules is 50 K. It is encouraging that the same value of
$T_{rot}$ applies to all three species. We discuss the implication
of this finding in Section \ref{sec:con}.

\section{Discussion and Conclusions}\label{sec:con}

The comparison in Section \ref{sec:comp} between the molecular
abundances obtained from the observations and from the models of
\citet{Baye10c} suggests that we can identify one model, Model E,
that produces results consistent with the observations. This model
has a low FUV radiation field, a low rate of heating by
dissipation, but a high cosmic ray ionization rate. Unpublished
data from the \citet{Baye10c} calculations show that cosmic ray
heating accounts for 85\% of all heating at either $A_{v}=$ 3 or 8
mag.

The enhanced cosmic ray flux may arise in a dynamical interaction,
as suggested by \citet{Ferl09}. Those authors also found that an
enhanced cosmic ray ionization rate was required to account for
the observations of the optical and infrared emitting components
of the filaments. It is interesting, therefore, that the results
from the present work also suggest that the regions of cluster gas
probed by millimetre and sub-millimetre emissions also require a
similarly enhanced cosmic ray ionization rate. A second inference
is that a high heating rate from energy dissipation is not
required; in fact, as Table \ref{tab:sum} indicates, a high
heating rate from sources other than cosmic rays appears to
inhibit a match between models and observations.

In the observations, we have detected only one line in each of
three molecular species, so a reliable analysis of the
observational data cannot be made. It would be useful for the
further study of galaxy cluster gas to make confirmed detections
of several lines of each of the three molecular tracers used in
this work. In the modelling of the chemistry of the gas in and
around NGC 1275 (\citealt{Baye10c}), some important physical
parameters (e.g. gas:dust ratio and metallicity) are poorly known
and undoubtedly have an impact on the model predictions.
Nevertheless, the present work shows the value of molecular line
observations of cluster gas, and it is interesting that the
results reported here for the molecular gas are reasonably
consistent with the results for regions probed by the optical and
infrared. The main conclusions of this work are that a cosmic ray
ionization rate enhanced by a factor of at least one hundred in
the cluster gas is required to establish the observed molecular
abundances, and that a high heating rate from other causes appears
to mitigate against that chemistry. In these circumstances, the
HCO$^{+}$ emission is strong.

\begin{table*}
  \caption{Total column density estimates (in cm$^{-2}$) obtained using the formula
  described in Section \ref{sec:resu}, for various values of $T_{rot}$. Since the
  there are observed upper limits for both NGC 1275 and the filamentary region, the
  corresponding derived total column densities presented here are thus also to be considered
  as upper limits (symbol $<$ used). The values of $T_{rot}$ has been selected with respect
  to the available online values of the partition function $Q(T_{rot})$ and such as
  to be compatible with the range of kinetic temperature displayed by the models (see Table
  \ref{tab:mod}).}\label{tab:resu}
  \begin{tabular}{l c c c c c c c}
  \hline
  & $T_{rot}=2.725$ K & $T_{rot}= 5$ K & $T_{rot}= 9.375$ K & $T_{rot}= 18.75$ K
  & $T_{rot}= 37.5$ K & $T_{rot}= 50$ K & $T_{rot}= 75$ K\\
  \hline
  NGC 1275 &&&&&&&\\
  \hline
  HCO$^{+}$ & $6.93 \times 10^{14}$ & $1.84 \times 10^{13}$ & $2.62 \times 10^{12}$
  & $9.45 \times 10^{11}$ & $6.22 \times 10^{11}$ & $5.86 \times 10^{11}$
  & $5.46 \times 10^{11}$\\
  CN & $2.42 \times 10^{14}$ & $1.91 \times 10^{13}$ & $5.02 \times 10^{12}$
  & $2.53 \times 10^{12}$ & $1.93 \times 10^{12}$ & $1.86 \times 10^{12}$
  & $1.79 \times 10^{12}$\\
  C$_{2}$H & $<9.03 \times 10^{15}$ & $<1.55 \times 10^{14}$ & $<1.74 \times 10^{13}$
  & $<5.40 \times 10^{12}$ & $<3.23 \times 10^{12}$ & $<2.94 \times 10^{12}$
  & $<2.66 \times 10^{12}$\\
  \hline
  \hline
  Filament &&&&&&&\\
  \hline
  HCO$^{+}$ & $<2.10 \times 10^{14}$ & $<5.58 \times 10^{12}$ & $<7.94 \times 10^{11}$
  & $<2.87 \times 10^{11}$ & $<1.88 \times 10^{11}$ & $<1.78 \times 10^{11}$
  & $<1.66 \times 10^{11}$\\
  CN & $<4.68 \times 10^{14}$ & $<3.69 \times 10^{13}$ & $<9.70 \times 10^{12}$
  & $<4.89 \times 10^{12}$ & $<3.73 \times 10^{12}$ & $<3.60 \times 10^{12}$
  & $<3.36 \times 10^{12}$\\
  C$_{2}$H & $<1.34 \times 10^{16}$ & $<2.52 \times 10^{14}$ & $<2.97 \times 10^{13}$
  & $<9.47 \times 10^{12}$ & $<5.74 \times 10^{12}$ & $<5.25 \times 10^{12}$
  & $<4.77 \times 10^{12}$\\
  \hline
  \end{tabular}
\end{table*}

\begin{table*}
  \caption{Fractional abundances (with respect to the total number of hydrogen nuclei)
  obtained using hypothesis described in Section \ref{sec:resu}, for various values of
  $T_{rot}$ and for $A_{v}=3$ mag and 8 mag (consistently with model
  predictions from \citealt{Baye10c}. See text in Section \ref{sec:comp}). See caption
  of Table \ref{sec:resu}.}\label{tab:abun}
  \begin{tabular}{l c c c c c c c}
  \hline
  & $T_{rot}=2.725$ K & $T_{rot}= 5$ K & $T_{rot}= 9.375$ K & $T_{rot}= 18.75$ K
  & $T_{rot}= 37.5$ K & $T_{rot}= 50$ K & $T_{rot}= 75$ K\\
  \hline
  NGC 1275 &&&&&&&\\
  \hline
  $A_{v}=3$ mag &&&&&&&\\
  HCO$^{+}$ & $1.44 \times 10^{-7}$ & $3.84 \times 10^{-9}$ & $5.45 \times 10^{-10}$
  & $1.97 \times 10^{-10}$ & $1.30 \times 10^{-10}$ & $1.22 \times 10^{-10}$
  & $1.14 \times 10^{-10}$\\
  CN & $5.04 \times 10^{-8}$ & $3.97 \times 10^{-9}$ & $1.05 \times 10^{-9}$
  & $5.27 \times 10^{-10}$ & $4.01 \times 10^{-10}$ & $3.87 \times 10^{-10}$
  & $3.72 \times 10^{-10}$\\
  C$_{2}$H & $<1.88 \times 10^{-6}$ & $<3.23 \times 10^{-8}$ & $<3.62 \times 10^{-9}$
  & $<1.12 \times 10^{-9}$ & $<6.72 \times 10^{-10}$ & $<6.12 \times 10^{-10}$
  & $<5.54 \times 10^{-10}$\\
  \hline
  $A_{v}=8$ mag &&&&&&&\\
  HCO$^{+}$ & $5.42 \times 10^{-8}$ & $1.44 \times 10^{-9}$ & $2.05 \times 10^{-10}$
  & $7.39 \times 10^{-11}$ & $4.86 \times 10^{-11}$ & $4.58 \times 10^{-11}$
  & $4.27 \times 10^{-11}$\\
  CN & $1.89 \times 10^{-8}$ & $1.49 \times 10^{-9}$ & $3.92 \times 10^{-10}$
  & $1.98 \times 10^{-10}$ & $1.51 \times 10^{-10}$ & $1.45 \times 10^{-10}$
  & $1.40 \times 10^{-10}$\\
  C$_{2}$H & $<7.06 \times 10^{-7}$ & $<1.21 \times 10^{-8}$ & $<1.36 \times 10^{-9}$
  & $<4.22 \times 10^{-10}$ & $<2.52 \times 10^{-10}$ & $<2.29 \times 10^{-10}$
  & $<2.08 \times 10^{-10}$\\
  \hline
  \hline
  Filament &&&&&&&\\
  \hline
  $A_{v}=3$ mag &&&&&&&\\
  HCO$^{+}$ & $<4.38 \times 10^{-8}$ & $<1.16 \times 10^{-9}$ & $<1.65 \times 10^{-10}$
  & $<5.97 \times 10^{-11}$ & $<3.93 \times 10^{-11}$ & $<3.70 \times 10^{-11}$
  & $<3.45 \times 10^{-11}$\\
  CN & $<9.74\times 10^{-8}$ & $<7.68 \times 10^{-9}$ & $<2.02 \times 10^{-9}$
  & $< 1.02 \times 10^{-9}$ & $<7.76 \times 10^{10}$ & $<7.50 \times 10^{-10}$
  & $<7.00 \times 10^{-10}$\\
  C$_{2}$H & $<2.79 \times 10^{-6}$ & $<5.24 \times 10^{-8}$ & $<6.18 \times 10^{-9}$
  & $<1.97 \times 10^{-9}$ & $<1.20 \times 10^{-9}$ & $<1.09 \times 10^{-9}$
  & $<9.93 \times 10^{-10}$\\
  \hline
  $A_{v}=8$ mag &&&&&&&\\
  HCO$^{+}$ & $<1.64 \times 10^{-8}$ & $<4.36 \times 10^{-10}$ & $<6.20 \times 10^{-11}$
  & $<2.24 \times 10^{-11}$ & $<1.47 \times 10^{-11}$ & $<1.39 \times 10^{-11}$
  & $<1.29 \times 10^{-11}$\\
  CN & $<3.65 \times 10^{-8}$ & $< 2.88 \times 10^{-9}$ & $< 7.58 \times 10^{-10}$
  & $<3.82 \times 10^{-10}$ & $<2.91 \times 10^{-10}$ & $<2.81 \times 10^{-10}$
  & $<2.62 \times 10^{-10}$\\
  C$_{2}$H & $<1.05 \times 10^{-6}$ & $<1.97 \times 10^{-8}$ & $<2.32 \times 10^{-9}$
  & $<7.40 \times 10^{-10}$ & $<4.49 \times 10^{-10}$ & $<4.10 \times 10^{-10}$
  & $<3.72 \times 10^{-10}$\\
  \hline
  \end{tabular}
\end{table*}

\begin{table*}
  \caption{Predicted values of the fractional abundances (with respect to the total
  number of hydrogen nuclei) of CN, C$_{2}$H and HCO$^{+}$ from
  models seen in \citet{Baye10c}. More precisely, Models A - E listed here correspond
  to Models 16, 14, 5, 18, and 12, respectively, in \citet{Baye10c} hence to models
  having high values of both $H$ and $\zeta$ with a low value of $I$; low values of $H$,
  $\zeta$ and $I$; low values of $H$ and $\zeta$ with a high value of $I$; high value
  of $H$ and low values of $\zeta$ and $I$ and low value of $H$, high value of $\zeta$ and
  low value of $I$, respectively. See the text in Section \ref{sec:comp}.}\label{tab:mod}
  \begin{tabular}{l c c c c c c c c}
  \hline
  Model & $A_{v}$ & $T_{K}$ & $H$ & $\zeta$ & $I$ & Frac. abun. & Frac. abun. & Frac. abun.\\
  & (mag) & (K) & (ergcm$^{-3}$ s$^{-1}$) & (s$^{-1}$) & (Habing) & CN & C$_{2}$H & HCO$^{+}$\\
  \hline
  A & 3 & 43.8 & $1.00 \times 10^{-20}$ & $5.00 \times 10^{-15}$ & 10 & $7.33 \times 10^{-9}$ & $3.32 \times 10^{-10}$
  & $4.69 \times 10^{-10}$\\
  A & 8 & 47.5 & $1.00 \times 10^{-20}$ & $5.00 \times 10^{-15}$ & 10 & $7.24 \times 10^{-9}$ & $2.74 \times 10^{-10}$
  & $5.94 \times 10^{-10}$\\
  \hline
  B & 3 & 13.8 & $1.00 \times 10^{-22}$ & $5.00 \times 10^{-17}$ & 10 & $1.74 \times 10^{-9}$ & $9.01 \times 10^{-12}$
  & $1.36 \times 10^{-11}$\\
  B & 8 & 14.9 & $1.00 \times 10^{-22}$ & $5.00 \times 10^{-17}$ & 10 & $1.81 \times 10^{-10}$ & $5.22 \times 10^{-12}$
  & $5.32 \times 10^{-11}$\\
  \hline
  C & 3 & 25.3 & $1.00 \times 10^{-22}$ & $5.00 \times 10^{-17}$ & 100 & $1.55 \times 10^{-9}$ & $5.50 \times 10^{-11}$
  & $1.31 \times 10^{-11}$\\
  C & 8 & 14.5 & $1.00 \times 10^{-22}$ & $5.00 \times 10^{-17}$ & 100 & $1.82 \times 10^{-10}$ & $5.28 \times 10^{-12}$
  & $5.16 \times 10^{-11}$\\
  \hline
  D & 3 & 74.8 & $1.00 \times 10^{-20}$ & $5.00 \times 10^{-17}$ & 10 & $2.51 \times 10^{-9}$ & $8.82 \times 10^{-11}$
  & $1.65 \times 10^{-10}$\\
  D & 8 & 63.3 & $1.00 \times 10^{-20}$ & $5.00 \times 10^{-17}$ & 10 & $8.61 \times 10^{-10}$ & $5.43 \times 10^{-12}$
  & $2.62 \times 10^{-10}$\\
  \hline
  E & 3 & 51.0 & $1.00 \times 10^{-22}$ & $5.00 \times 10^{-15}$ & 10 & $1.82 \times 10^{-9}$ & $2.68 \times 10^{-10}$
  & $1.37 \times 10^{-10}$\\
  E & 8 & 60.2 & $1.00 \times 10^{-22}$ & $5.00 \times 10^{-15}$ & 10 & $1.99 \times 10^{-9}$ & $2.95 \times 10^{-10}$
  & $2.94 \times 10^{-10}$\\
  \hline
  \end{tabular}
\end{table*}

\begin{table*}
\caption{Summary of the conclusions obtained when comparing the
estimates of the fractional abundances derived from the
observations to the model predictions (see Section
\ref{sec:comp}). We have assumed that models and observations are
in agreement (symbol `+' used) when their values differ from less
than or equal to a factor of 5 difference. Otherwise, the symbol
`-' is used. `(3)' or `(8)' means that the fractional abundances
at $A_{v}=3$ mag or at $A_{v}=8$ mag are well reproduced by the
models (but not at both $A_{v}$s), respectively.}\label{tab:sum}
\begin{tabular} {l c c c c c c}
\hline
    Model  &        C$_{2}$H  & C$_{2}$H & CN & CN & HCO$^{+}$ & HCO$^{+}$\\
     Parameters & centre & filament & centre & filament & centre & filament\\
\hline
Model A: $I$ low, $H$ high, $\zeta$ high& $+$ & $+$ & -   & -   & (3) & -  \\
Model B: $I$ low, $H$ low, $\zeta$ low  & -   & -   & $+$ & $+$ & (8) & (8)\\
Model C: $I$ high, $H$ low, $\zeta$ low &  -  &  -  & $+$ & $+$ & (8) & (8)\\
Model D: $I$ low, $H$ high, $\zeta$ low & -   & -   & $+$ & $+$ & (3) & (3)\\
Model E: $I$ low, $H$ low, $\zeta$ high & $+$ & $+$ & (3) & (3) & (3) & (3)\\
\hline
\end{tabular}
\end{table*}

\section*{Acknowledgments}

EB, TWH and SV acknowledges financial support from STFC. We thank
Alastair Edge for his constructive referee's report, which helped
us improve the paper.

\newcommand{\apj}[1]{ApJ, }
\newcommand{\apss}[1]{Ap\&SS, }
\newcommand{\aj}[1]{Aj, }
\newcommand{\apjs}[1]{ApJS, }
\newcommand{\apjl}[1]{ApJ Letter, }
\newcommand{\aap}[1]{A\&A, }
\newcommand{\aaps}[1]{A\&A Suppl. Series, }
\newcommand{\araa}[1]{Annu. Rev. A\&A, }
\newcommand{\aaas}[1]{A\&AS, }
\newcommand{\bain}[1]{Bul. of the Astron. Inst. of the Netherland,}
\newcommand{\mnras}[1]{MNRAS, }
\newcommand{\nat}[1]{Nature, }
\newcommand{\araaa}[1]{ARA\&A, }
\newcommand{\planss}[1]{Planet Space Sci., }
\newcommand{\jrasc}[1]{Jr\&sci, }
\newcommand{\pasj}[1]{PASJ, }
\newcommand{\pasp}[1]{PASP, }

\bibliographystyle{mn2e_2}
\bibliography{references}

\bsp

\label{lastpage}

\end{document}